\documentclass[journal]{IEEEtran}
\usepackage{ifpdf}
\usepackage{gensymb}

\usepackage{cite}

\usepackage{enumitem}
\usepackage{stfloats}

\ifCLASSINFOpdf
   \usepackage[pdftex]{graphicx}
\else
   \usepackage[dvips]{graphicx}
\fi

\usepackage{amsmath}
\usepackage{algorithmic}

\usepackage{array}

\usepackage[tight,footnotesize]{}
\usepackage{subfigure}

\usepackage[utf8]{inputenc}
\usepackage[T1]{fontenc}

\usepackage{url}

\begin{document}
%
\title{Linear Hybrid Asymmetrical Load-Modulated Balanced Amplifier with Multi-Band Reconfigurability and Antenna-VSWR Resilience}

%
%
%

\author{Jiachen~Guo,~\IEEEmembership{Student Member,~IEEE,}
       Yuchen~Cao,~\IEEEmembership{Member,~IEEE,}
        and Kenle~Chen,~\IEEEmembership{Senior Member,~IEEE}

\thanks{“© 2024 IEEE.  Personal use of this material is permitted.  Permission from IEEE must be obtained for all other uses, in any current or future media, including reprinting/republishing this material for advertising or promotional purposes, creating new collective works, for resale or redistribution to servers or lists, or reuse of any copyrighted component of this work in other works.”}

\thanks{This work was supported in part by the National Science Foundation under Award No.~2218808.}

\thanks{Color versions of one or more of the figures in this paper are available online at https://ieeexplore.ieee.org.
}
\thanks{Digital Object Identifier 10.1109/TMTT.2024.3381845
}
}


%
%

\markboth{IEEE Trans.~Microwave Theory and Techniques}%
{Shell \MakeLowercase{\textit{et al.}}: Bare Demo of IEEEtran.cls for Journals}
%



\maketitle

\begin{abstract}

This paper presents the first-ever highly linear and load-insensitive three-way load-modulation power amplifier (PA) based on reconfigurable hybrid asymmetrical load modulated balanced amplifier (H-ALMBA). Through proper amplitude and phase controls, the carrier, control amplifier (CA), and two peaking balanced amplifiers (BA1 and BA2) can form a linear high-order load modulation over wide bandwidth. Moreover, it is theoretically unveiled that the load modulation behavior of H-ALMBA can be insensitive to load mismatch by leveraging bias reconfiguration and the intrinsic load-insensitivity of balanced topology. Specifically, the PA's linearity and efficiency profiles can be maintained against arbitrary load mismatch through $Z_\mathrm{L}$-dependent reconfiguration of CA supply voltage ($V_\mathrm{DD,CA}$) and turning-on sequence of BA1 and BA2. Based on the proposed theory, an RF-input linear H-ALMBA is developed with GaN transistors and wideband quadrature hybrids. Over the design bandwidth from $1.7$-$2.9$ GHz, an efficiency of $56.8\%$$-$$72.9\%$ at peak power and  $49.8\%$$-$$61.2\%$ at $10$-dB PBO are measured together with linear AMAM and AMPM responses. In modulated evaluation with 4G LTE signal, an EVM of $3.1\%$, ACPR of $-39$ dB, and average efficiency of up to $52\%$ are measured. Moreover, the reconfigurable H-ALMBA experimentally maintains an excellent average efficiency and linearity against arbitrary load mismatch at $2:1$ VSWR, and this mismatch-resilient operation can be achieved at any in-band frequencies. The overall measured performance favorably outperforms the state-of-the-art.


\end{abstract}

\begin{IEEEkeywords}
Balanced amplifier, Doherty power amplifier, linearity, load mismatch, load modulation, reconfigurable.
\end{IEEEkeywords}

%
\IEEEpeerreviewmaketitle

\section{Introduction}
%
%
%
%
\begin{figure}
\centering
\includegraphics[width=88mm]{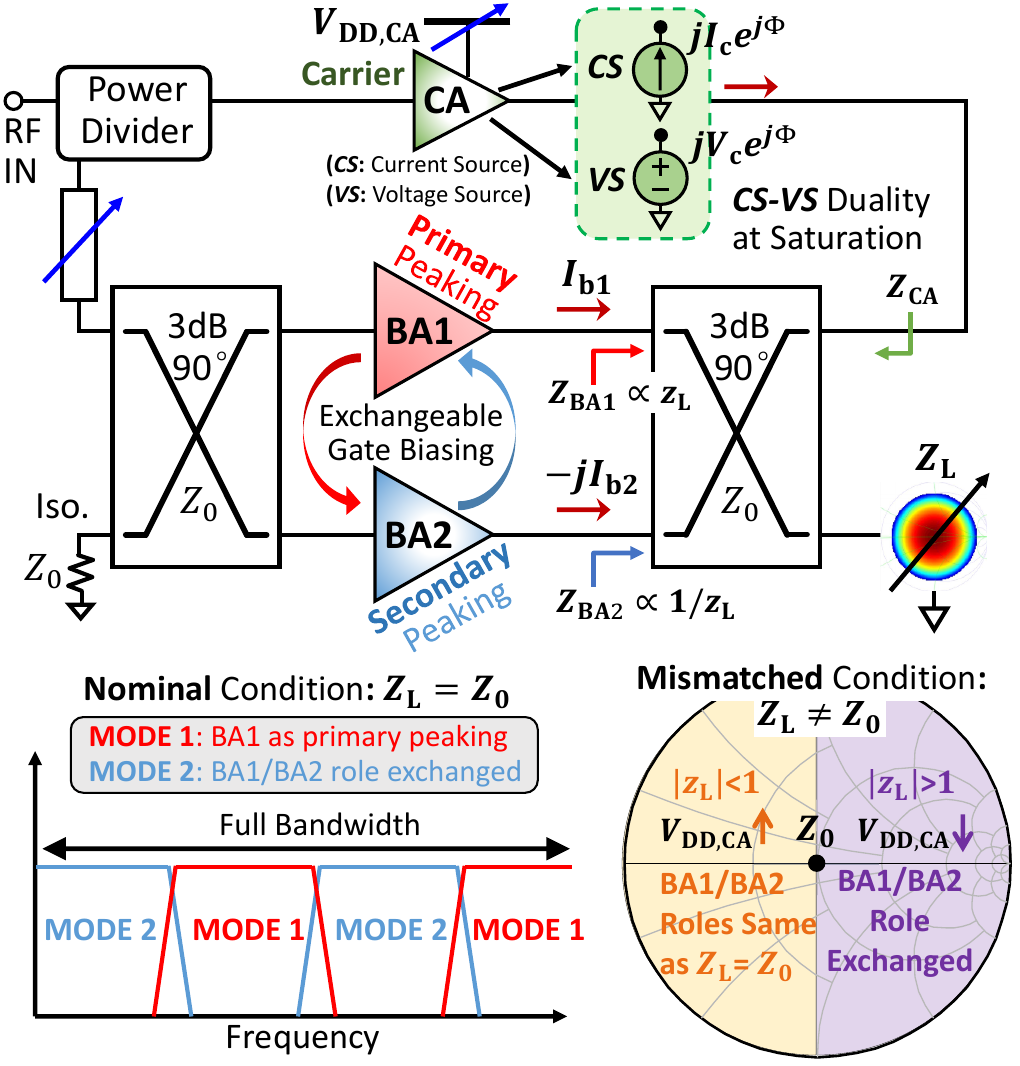}
\vspace{-20pt}
\caption{Overview of H-ALMBA with varactor-less reconfiguration for wide bandwidth and load insensitiveness.}
\label{fig1:OVERVIEW}
\vspace{-18pt}
\end{figure}

\IEEEPARstart{H}{IGHLY} spectrally efficient modulation techniques for high-data-rate transmissions are increasingly prevalent in new communication standards, catering to the rising demand for faster wireless data transfer. However, these sophisticated modulation methods significantly elevate the Peak-to-Average Power Ratio (PAPR) of signals. This increase is primarily due to the use of numerous sub-carriers and advanced higher-order digital mapping techniques, such as 4096QAM. Effectively amplifying these high-PAPR signals necessitates power amplifiers (PAs) with enhanced efficiency across a wide output back-off (OBO) range. Furthermore, as wireless standards evolve and the spectrum extends into higher frequencies, such as the $3-5$ GHz range newly allocated for 5G, PAs are required to accommodate wider bandwidths and also support multimode and multiband operations, thereby ensuring that the system remains manageable in terms of complexity and cost. As a result, enhanced back-off efficiency range and broad bandwidth gradually become crucial targets for contemporary PAs. In addition to established technologies like Envelope Tracking (ET) \cite{ET} and Doherty Power Amplifiers \cite{doherty1,doherty2,doherty3}, recent advancements have introduced innovative architectures and techniques \cite{new_architecture2,new_architecture3}, such as load-modulated balanced amplifiers \cite{PDLMBA,LMBA_MWCL2016,DLMBA_TMTT2018,digital_lmba,LMBA_chappello,LMBA_chappidi,LMBA_vivien,WBLMBA_TMTT2017_TB,Cripps2018,Broadband_cm,ALMBA}. These new developments offer remarkable improvements in efficiency, Output Back-Off (OBO) range, and bandwidth capabilities, representing significant strides in amplifier technology.

On the other hand, massive MIMO has been a cornerstone of 5G systems for enhanced user capacity through spatial diversity \cite{MIMObook}. However, the deployment of large, densely-packed antenna arrays introduces a notable challenge: strong mutual coupling between the individual radiating elements \cite{coupling_effect}. This leads to a classical issue of antenna scan impedance, which can vary over a substantial range during beam steering \cite{Magazine_Fager_5GTX}. Recent research indicates that active antenna impedance mismatch can reach a voltage standing wave ratio (VSWR) of up to $6:1$ during beam scanning \cite{chen2015effect,Mutual_coupling_MIMO}.  As the immediate stage preceding the antenna, PAs in the array are exposed to constantly shifting loads. This can lead to notable performance variations in their performance, impacting linearity, stability, output power, and efficiency. To address the challenge of load mismatch in PAs within array systems, various solutions have been developed until now. Traditionally, a circulator or isolator is positioned between the PA and the antenna load to mitigate interactions between them. However, the common ferrite-based isolators present drawbacks: they are costly, non-integrable with other system components, and tend to be bulky, which limits their practicality in compact and integrated massive-scale deployment. Alternatively, incorporating a tunable matching network between the Power Amplifier (PA) and the antenna offers a dynamic solution to compensate for variations in antenna impedance mismatch \cite{TMN_1,TMN_2,TMN_3,TMN_4}. This approach allows for real-time adjustments, ensuring more efficient and stable PA performance in response to changing load conditions. 

Note that all of the solutions mentioned above necessitate the incorporation of additional building blocks, leading to complication at the system level with increased losses, costs, and physical size. Therefore, there is a compelling interest in integrating tuning into the PA stage, and multiple reconfiguration-based load-insensitive PA topologies have been reported recently. In \cite{wanhua_jssc,Lyu_B2D_2020,LinearQB-DPA,QB_dpa_2022}, several quadrature-coupler-based reconfigurable Doherty PA (DPA) are presented by utilizing exchangeable gate biasing, adaptive power splitting ratio, and tunable carrier-peaking phase offset to configure multiple load-dependent DPA operational modes, so as to compensate for the performance degradation at load mismatch. A comparable approach is reported in \cite{Multiport_Chappidi} that presents a generic representation of multi-port load-modulation combiner. This design is notable for its tolerance to load mismatch, expanded bandwidth, and a reconfigurable Output Back-Off (OBO) range. Additionally, a plethora of load-insensitive DPAs are introduced in \cite{dynamivCA_ref,Self-Healing}, which utilizes variable DC supply voltage and multi-input phase tuning to sustain the PA performance fluctuating load impedance conditions. In summary, it is important to note that the existing solutions are mostly tackling two-way load modulation, which has limited OBO range and back-off efficiency. 

Derived from PD-LMBA \cite{PDLMBA} as well as other Doherty-like LMBAs \cite{DLMBA_TMTT2018} that are typically based on a two-way modulation, the novel H-ALMBA architecture enables three-way load modulation by separately turning on the transistors in balanced amplifier \cite{CM_LMBA,Lyu_HLMBA,guo_lmdba}. This new mode not only achieves three efficiency peaks throughout a larger OBO range but also eliminates the undesired efficiency drop in the middle of the back-off levels. Compared to the three-way DPA with a general difficulty for wideband design, H-ALMBA also perfectly inherits the wideband nature from PD-LMBA, while offering an enhanced reconfigurability for load-mismatch-resilient operation, offering a substantial improvement in both performance (not just efficiency as the reconfiburable PD-LMBA in \cite{1d-pdlmba}) and design versatility. 

A proof-of-concept demonstration of the reconfigurable H-ALMBA is preliminarily presented in our IMS paper \cite{guoIMS2023}, which is significantly expanded in the follow aspects: 1) We unveil the very first theoretical analysis of the linear H-ALMBA mode that is developed from the original three-way H-ALMBA \cite{CM_LMBA}, as well as its capability of counteracting load mismatch through intrinsic circuit reconfiguration; 2) As compared to \cite{guoIMS2023}, circuit simulation using emulated model is comprehensively conducted to verify the theory; 3) Further optimized and conclusive measurement results are experimentally presented. Specifically, the theory is initially established by analytically formulating the linear H-ALMBA using piece-wise linear models of BA1, BA2 and CA, encompassing different operational modes in both low-power and high-power load-modulation regions. By setting asymmetrical turning-on thresholds of BA1 and BA2 together with linearity-enhanced phase control, the H-ALMBA's AMAM and AMPM performance can be linearized with a slight compromise of efficiency. Next, a generalized theory of mismatch-resilient H-ALMBA is analytically established, which is based on a unique technique of characterizing CA as a duality between current source (CS) and voltage source (VS) \cite{NiteeshIMS2022,1d-pdlmba}. The derived theory indicates that the efficiency and linearity profiles of H-ALMBA can be effectively sustained by performing load-dependent reconfiguration, i.e., exchanging the gate biasing of BA1 and BA2 and adjusting the supply voltage of CA according to the magnitude of load impedance, $|Z_\mathrm{L}|$, as illustrated in Fig.~\ref{fig1:OVERVIEW}. The same type of reconfiguration can also be applied to achieve optimal H-ALMBA performance at different frequency bands. The developed experimental prototype of this new linear load-insensitive LM architecture significantly outperforms state-of-the-art for both nominal matched load conditions and a $2:1$ VSWR. Such a successful demonstration highlights its substantial potential for integration into array-based massive MIMO systems.

\begin{figure}
\centering
\includegraphics[width=73mm]{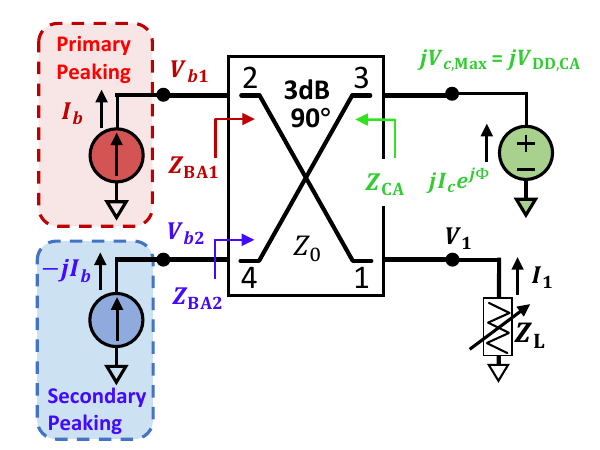}
\vspace{-5pt}
\caption{Generalized schematic of H-ALMBA consist of peaking and control amplifiers.}
\label{fig2:HALMBA_schematic}
\vspace{-15pt}
\end{figure}


\section{Theoretical Analysis of Highly Linear H-ALMBA with Load Insensitivity}
\label{sec:Theory_Norm}

 The load-modulation of H-ALMBA can be modeled as three excitation sources driving the output quadrature coupler as shown in Fig.~\ref{fig2:HALMBA_schematic}, and it can be analytically described using the impedance matrix introduced in the following matrix as
\begin{equation}
 \begin{bmatrix}
   V_{1} \\
   V_{2} \\
    V_{3}\\
    V_{4} 
  \end{bmatrix}
=
Z_{\mathrm{0}}
 \begin{bmatrix}
  0&0&+j&-j\sqrt{2}\\
    0&0&-j\sqrt{2}&+j\\
   +j&-j\sqrt{2}&0&0 \\
   -j\sqrt{2}&+j&0&0
  \end{bmatrix}
  \begin{bmatrix}
   I_{1} \\
   I_{2} \\
    I_{3}\\
    I_{4} 
  \end{bmatrix}
  \label{eq:Matrix_Norm}
\end{equation}
where $V_1=-I_{1}/Z_{0}$ stands for the output terminal, $I_{2}=I_{b1}$ and $I_{4}=-jI_{b2}$ represent the input RF currents from BA1 and BA2, and $I_{3}=jI_{c}e^{j\phi}$ denotes the CA current. Using the matrix above, the load impedances of BA1, BA2 and CA under matched condition can be calculated as \cite{ALMBA}
\begin{align}
& Z_\mathrm{BA1}=Z_\mathrm{0}(\frac{\sqrt{2}I_{c}e^{j\phi}}{I_{b1}}+\frac{I_{b2}}{I_{b1}});\label{eq:Z_BA1_AS}
\end{align}

\begin{align}
&Z_\mathrm{BA2}=Z_\mathrm{0}(2+\frac{\sqrt{2}I_{c}e^{j\phi}}{I_{b2}}-\frac{I_{b1}}{I_{b2}});
\label{eq:Z_BA2_AS}
\end{align}

\begin{align}
& Z_\mathrm{CA}=Z_\mathrm{0}(1+\frac{\sqrt{2}(I_{b2}-I_{b1})}{I_{c}e^{j\phi}}).
\label{eq:Z_CA_AS}
\end{align}
The above equations provide a comprehensive explanation of the generalized quadrature-coupled load modulation behavior, which encompasses the original LMBA \cite{LMBA_IMS2017_Cripps} and all possible variations. It is clear that the modulation of $Z_\mathrm{BA1}$ and $Z_\mathrm{BA2}$ can be controlled by altering the amplitude and phase of $I_\mathrm{c}$. Simultaneously, the load on the carrier amplifier, $Z_\mathrm{CA}$, is determined by the difference between $I_\mathrm{b1}$ and $I_\mathrm{b2}$. H-ALMBA takes advantage of different turn-on sequences of BA1 and BA2, which not only allows for the adjustment of $I_\mathrm{b1}$ and $I_\mathrm{b2}$ at various OBO levels but also enables a three-way load modulation behavior.

\vspace{-12pt}
\subsection{Modeling of Carrier and Peaking Generators}
\label{subsec:CG model}

\begin{figure}
\centering
\includegraphics[width=80mm]{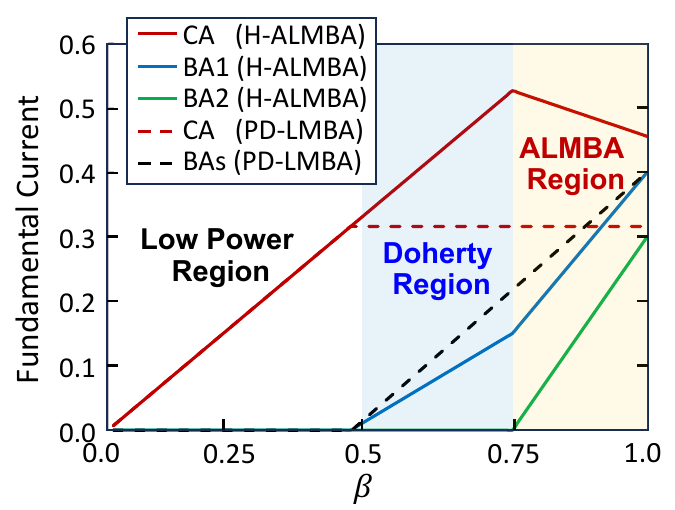}
\vspace{-13pt}
\caption{Comparison of normalized fundamental currents between H-ALMBA and PD-LMBA modes.}
\label{fig3:theory_current}
\vspace{-17pt}
\end{figure}

\begin{figure}
\centering
\includegraphics[width=80mm]{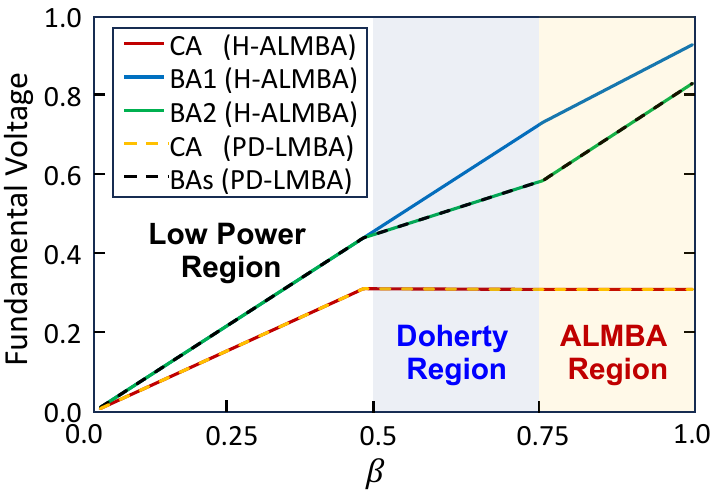}
\vspace{-13pt}
\caption{Comparison of normalized fundamental voltages of each path in
H-ALMBA and PD-LMBA mode}
\label{fig4:theroy_voltage}
\vspace{-17pt}
\end{figure}
As derived from H-ALMBA theory \cite{lMBA6}, with different turning-on sequences of CA, BA1 and BA2, a three-way Doherty-like PA is realized, and its dynamic operation can be divided into three different power regions as Low-Power (CA standalone), Doherty (CA+BA1), and ALMBA (CA+BA1+BA2) regions. To theoretically analyze the load modulation behavior of three sub-amplifiers from H-AMLBA, the transistors' current models are established respectively in these three different regions\cite{Crippsbook,SLMBA,CM_LMBA}. Additionally, the drain current is defined only with the magnitude information at this stage, and the phase relationship will be considered
when analyzing the load-modulation behaviors.

\begin{figure}[t]
\centering
\includegraphics[width=80mm]{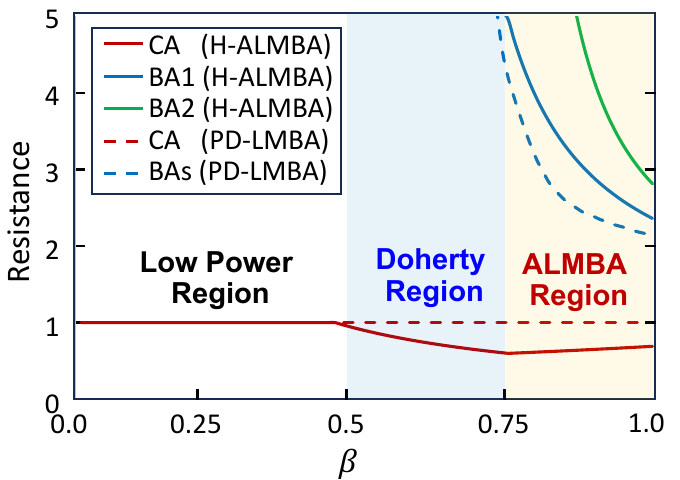}
\vspace{-13pt}
\caption{Comparison of load resistance between three sub-amplifiers in H-ALMBA and PD-LMBA.}
\label{fig5:theory_resistance}
\vspace{-25pt}
\end{figure}

\begin{enumerate}
    \item \textbf{\textit{Low-Power Region} ($0<\beta<\beta_\mathrm{lbo}$)}:
In this region with the carrier amplifier CA solely operation, its current, $i_\mathrm{ca,lp}$ can be expressed using the piece-wise linear model of Class-B mode:
 \begin{equation}
i_\mathrm{ca,lp}(\beta) =\left\{
 \begin{aligned}
&\beta I_\mathrm{Max,C}\cdot\cos{\theta}, &&-\frac{\pi}{2}\leq\theta < \frac{\pi}{2}\\
 &0 &&\text{otherwise}
 \end{aligned}
 \right.
\end{equation}
where $\beta$ stands for the normalized magnitude of the input voltage driving level, and $I_\mathrm{Max,C}$ is the maximum channel current of the transistor device. Its fundamental and DC current can be derived using Fourier transform as\begin{align}
     i_\mathrm{ca,lp}[0] = \frac{\beta I_\mathrm{Max,C}}{\pi};~~~~\\
   ~~ i_\mathrm{ca,lp}[1] = \frac{\beta I_\mathrm{Max,C}}{2}=I_c.
\end{align}
In this case, according to \eqref{eq:Z_CA_AS}, the impedance of CA is equal to $Z_\mathrm{0}$ which is also plotted in Fig.~\ref{fig5:theory_resistance}. The impedance of BA1 and BA2 is close to infinite as shown in Fig.~\ref{fig5:theory_resistance} since they are not turned on as an open circuit.

\item \textbf{\textit{Doherty Region} ($\beta_\mathrm{lbo}\leq\beta< \beta_\mathrm{hbo}$)}:
When the driving level reaches lower back-off (LBO), the CA reaches to its voltage saturation forming the first efficiency peak, and at the same time the primary peaking amplifier (BA1) is turned on to cooperate with CA as a Doherty PA. In the case of symmetrical PD-LMBA\cite{PDLMBA}, the CA is not load-modulated as peaking amplifier (BA) turns on, and it is thus over-driven by the continuously increased input power and remains saturated, as the red dash line illustrated in Fig.~\ref{fig3:theory_current}. While in the Doherty region of H-ALMBA model, CA current is modeled in the same way as the main amplifier of standard Doherty PA, which continues increasing linearly with $\beta$ at the same slope of low-power region until the secondary peaking ampifier (BA2) turns on at $\beta_\mathrm{hbo}$:
\begin{equation}
I_c=i_\mathrm{ca,doherty}(\beta)[1]=\frac{\beta I_\mathrm{Max,C}}{2}, ~~~~~\beta_\mathrm{lbo}\leq\beta\leq \beta_\mathrm{hbo}
\label{eq:I_CA_doherty}
\end{equation}
The voltage of BA1 in Doherty region can be derived from \eqref{eq:Matrix_Norm} as:
\begin{align}
& V_\mathrm{BA1,doherty}=Z_\mathrm{0}j(I_{b2}-\sqrt{2}I_{c})=-Z_\mathrm{0}j\sqrt{2}I_{c}.
\label{eq:V_BA1_doherty}
\end{align}

Likewise, The voltage of CA can be denmonstrated as:
\begin{align}
& V_\mathrm{CA,doherty}=Z_\mathrm{0}j(I_{c}-\sqrt{2}I_{b1}).
\label{eq:V_CA_doherty}
\end{align}

In order to keep the voltage saturation and solve the overdriving issue in Doherty region, the current of BA1 is no longer a single-slope curve during load modulation as in PD-LMBA indicated by the black dash line in Fig.~\ref{fig3:theory_current}. According to \eqref{eq:V_CA_doherty}, in order to maintain a constant $V_\mathrm{CA,doherty}$, the slopes of CA and BA1 fundamental currents with respect to $\beta$ should be constant. Thus, the scaled current curve of BA1 is plotted in Fig.~\ref{fig3:theory_current}, and it can be derived as \begin{equation}
\begin{aligned}
 &I_{b1}=i_\mathrm{ba1,doherty}(\beta)[1] =\\&
\frac{\sqrt{2}(\beta-\beta_\mathrm{lbo})}{4} I_\mathrm{Max,C}, &&\beta_\mathrm{lbo}\leq\beta\leq\beta_\mathrm{hbo}
\end{aligned}
\label{eq:I_BA_doherty}
\end{equation}
 
where $\beta_\mathrm{lbo}$ is the BA1 turning on point that is set to $0.5$ for instance in the theoretical analysis. With the determined CA and BA currents in \eqref{eq:I_CA_doherty} and \eqref{eq:I_BA_doherty}, the behavior of H-ALMBA in Doherty region is equivalent to that of a standard Doherty PA \cite{LinearQB-DPA}, so that the AM-AM and AM-PM of H-ALMBA can both be linear in this region. 

\item \textbf{\textit{ALMBA Region} ($\beta\geq\beta_\mathrm{hbo}$)}:
As the input driving level increases above the high back-off (HBO) into the ALMBA region, CA reaches full saturation at HBO and the current start to decrease due to the load impedance increases as BA2 turns on\cite{CM_LMBA}. Nevertheless, in this regions, CA remains voltage saturated such that its current source model can be transferred to a constant voltage source as shown in Fig.~\ref{fig4:theroy_voltage}, and the corresponding fundamental current of CA can be calculated as the following:\begin{equation}
 I_c=i_\mathrm{ca,almba}(\beta)[1] =\frac{2 V_\mathrm{DD,CA}}{Z_\mathrm{CA}(\beta)}.
 \label{Ica_almba}
\end{equation}
where $V_\mathrm{DD,CA}$ is the maximum fundamental voltage allowed for the power device of CA and $Z_\mathrm{CA}$ can be determined using (\ref{eq:Z_CA_AS}). Moreover, the peaking amplifiers, BA1 and BA2, are both turned on in ALMBA region, and the fundamental voltage of BA1 and BA2 in ALMBA region can be expressed as:
\begin{align}
& V_\mathrm{BA1,almba}=Z_\mathrm{0}j(I_{b2}-\sqrt{2}I_{c}).\label{eq:V_BA1_almba}
\\
& V_\mathrm{BA2,almba}=Z_\mathrm{0}(j(I_{b1}+\sqrt{2}I_{b2}-\sqrt{2}I_{c}).
\label{eq:V_BA2_almba}
\end{align}

The fundamental current of BA1 can be expressed by satisfying the following boundary condition:
\begin{equation}
i_\mathrm{ba1,doherty}(\beta_\mathrm{hbo})[1] =i_\mathrm{ba1,almba}
(\beta_\mathrm{hbo})[1]
\label{relation_boundary}
\end{equation}
where $\beta_\mathrm{hbo} = 0.75$ as a standard values of H-ALMBA, denoting the turning on power level BA2, such that the fundamental currents of BA1 and BA2 can be derived as

\begin{equation}
 \begin{aligned}
&I_{b1}=i_\mathrm{ba1,almba}(\beta)[1] =
\frac{(\beta-\beta_\mathrm{hbo})}{1-\beta_\mathrm{hbo}}\lambda I_\mathrm{Max,B}+\\&\frac{i_\mathrm{ba1,doherty}(\beta_\mathrm{hbo})[1](1-\beta)}{1-\beta_\mathrm{hbo}}, ~~~~~\beta_\mathrm{hbo}\leq\beta \leq 1
 \end{aligned}
 \label{eq:BA1_almba}
\end{equation}

 \begin{align}
 I_{b2}=i_\mathrm{ba2,almba}(\beta)[1] =
&\frac{\beta-\beta_\mathrm{hbo}}{1-\beta_\mathrm{hbo}} \gamma I_\mathrm{Max,B},~~~~~~~~\nonumber\\ 
&~~~~~~~~~~\beta_\mathrm{hbo}\leq\beta \leq 1
 \label{eq:BA2_almba}
\end{align}

where $I_\mathrm{Max,B}$ is the maximum current that the peaking device can provide at the peak-power level. Since BAs are biased in Class-C, their maximum fundamental current magnitudes are smaller than Class-B ($=0.5I_\mathrm{Max,B}$). Thus, two scalar factors, $\lambda$ and $\gamma$, are used in \eqref{eq:BA1_almba} and \eqref{eq:BA2_almba} to determine the normalized ratio with respect to the maximum current magnitude of Class-B. In this analysis, the values of $\lambda$ and $\gamma$ are set to $0.4$ and $0.3$, respectively, to represent the different conduction angles of BA1 and BA2 due to the different turning-on thresholds.

\end{enumerate}

\vspace{-12pt}
\subsection{Load Modulation Analysis and High-Linearity Design of H-ALMBA}
\label{subsec:CA_mismatch}
The detailed analysis of load-modulation behavior of H-ALMBA is performed and the load impedances of CA, BA1 and BA2 over the entire power region can be calculated using \eqref{eq:Z_BA1_AS}-\eqref{eq:Z_CA_AS}, which are expressed as
\begin{align}
    &Z_\mathrm{CA}=\frac{Z_\mathrm{0} V_\mathrm{DD,CA}}{V_\mathrm{DD,CA}+\sqrt{2}(I_{b1}-I_{b2})e^{-j\phi}Z_\mathrm{0}} \label{eq:ZCA}\\
    &Z_\mathrm{BA1}=2Z_\mathrm{0}+\frac{\sqrt{2} V_\mathrm{DD,CA}e^{j\phi}-Z_\mathrm{0}I_{b2}}{I_{b1}} \label{eq:ZBA1}\\
    &Z_\mathrm{BA2}=\frac{Z_\mathrm{0}I_{b1}+\sqrt{2} V_\mathrm{DD,CA}e^{j\phi}}{I_{b2}} \label{eq:ZBA2}
\end{align}
Please note that the above equations present an unified formulation of three sub-amplifiers' load impedances for the entire power region, where $I_{b1}$ and $I_{b2}$ in the equation are different at three power region as derived in Subsec.~\ref{subsec:CG model}. For example, in low-power region with $I_{b1}=I_{b2}=0$, $Z_\mathrm{CA}=Z_0$ can be derived from \eqref{eq:ZCA}. In Doherty region, $I_{b1} = i_\mathrm{ba1,doherty}[1]$ and $I_{b2} = 0$ such that the load impedance of CA only depends on the BA1 fundamental component and is modulated to a lower value as the red cure shown in the Fig.~\ref{fig5:theory_resistance}, which is mathematically calculated using the equations above. As BA2 is turned on in ALMBA region, $Z_\mathrm{CA}$ is reversely modulated due to the sharply increased $I_{b2}$ ($=i_\mathrm{ba2,almba}[1]$). The impedances of three amplifiers across the entire power range are illustrated in Fig.~\ref{fig5:theory_resistance}.

Moreover, from the $Z$-matrix in \eqref{eq:Matrix_Norm}, the output voltage $V_\mathrm{out}$ can be calculated as follow:\begin{equation}
       V_\mathrm{out}=V_\mathrm{1}= Z_\mathrm{0}(j I_{c}e^{j\phi} - j\sqrt2I_{b2}),
    \label{eq:theory_Vout}
\end{equation}where we can find that the output voltage is only dependent on the fundamental currents of CA and BA2 and the CA-BA phase offset $\phi$, which together determine the overall linearity of H-ALMBA in terms of AMAM and AMPM.
\begin{figure}
\centering
\includegraphics[width=80mm]{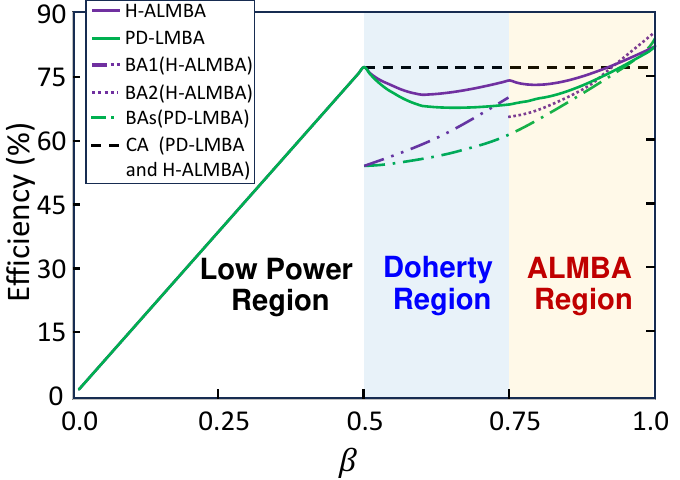}
\vspace{-10pt}
\caption{Comparison of efficiency between H-ALMBA and PD-LMBA.}
\label{fig:eff_comp}
\vspace{-15pt}
\end{figure}
In low-power and Doherty regions where $V\mathrm{out}$ is only dependent on CA current, H-ALMBA is intrinsically linear as AMAM and AMPM responses plotted in the Figs.~\ref{fig6:theory_swep_eff}(a) and (b). In ALMBA region, it can be seen that the gain shows an expansion behavior due to the extra contribution of BA2. Nevertheless, this gain expansion can be favorably utilized to compensate for the soft compression of three amplifiers, which are all driven towards full saturation in this ALMBA region. Overall, the AMAM of H-ALMBA can be fully linearized with optimal gain behaviors of three amplifiers, which can be realistically achieved with proper bias tuning. For AMPM in ALMBA region, as indicated by \eqref{eq:theory_Vout}, the red curve in Fig.~\ref{fig6:theory_swep_eff}(a) shows that the AMPM response can be ideally flattened when the BA-CA phase offset is set to $\phi=0$. Realistically, the AMPM is normally distorted due to the non-linear parasitic capacitors of the transistors, which can be effectively offset by forming a reverse AMPM behavior of load modulation that is physically realized through properly tuning the CA-BA phase offset, $\phi$. Thus, for different in band frequency, we can set an optimal value of $\phi$ to compensate for the AMPM distortion with a slightly degradation of efficiency as shown in Fig.~\ref{fig6:theory_swep_eff}(c).

\begin{figure}
\centering
\includegraphics[width=82
mm]{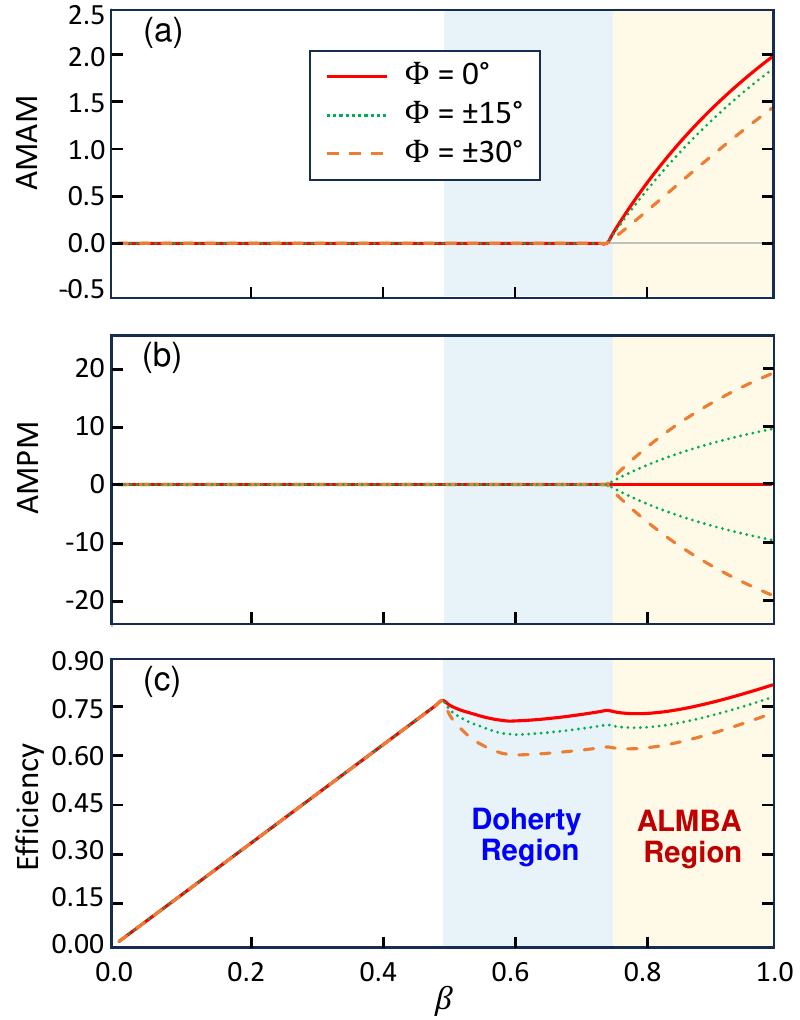}
\vspace{-10pt}
\caption{Performance of the proposed H-ALMBA versus different values of $\phi$: (a) AMAM, (b) AMPM, (c) Efficiency.}
\label{fig6:theory_swep_eff}
\vspace{-18pt}
\end{figure}

\vspace{-12pt}
\subsection{Reconfiguration for Resilience against Load Mismatch}
\label{subsec:theory_mismatch}
When the load is mismatched, the operational analysis of H-ALMBA is divided into the same three power regions respectively.
 \begin{enumerate}
    \item \textbf{\textit{Low-Power Region} ($P_\mathrm{OUT} < P_\mathrm{Max}/\mathrm{LBO}$)}: when the power level is below the LBO with BA turned off, BA1 and BA2 are open-circuited, and the associated quadrature coupler acts as an quarter-wave inverter ideally such that the corresponding load impedance, $Z_\mathrm{CA}$, is inverted by the quadrature coupler from $Z_\mathrm{L}$, given by \cite{1d-pdlmba}\begin{align}
     &z_\mathrm{CA,LP} = 1/z_\mathrm{L}=y_\mathrm{L}.
     \label{eq:Z_CA_LP}
     \end{align}The corresponding saturation power of CA, $P_\mathrm{CA,Sat.}$ at the mismatched load can be derived as:
     \begin{equation}
    P_\mathrm{CA,Sat.}=\frac{ V_\mathrm{DD,CA}^{2}}{2Z_\mathrm{0}}
\frac{\mathrm{real}(z_\mathrm{CA})}{\lvert z_\mathrm{CA} \rvert^2}.
\label{eq:pca2}
\end{equation} Accordingly, based on the loadline theory, the value of  $V_\mathrm{DD,CA}$ should ideally scale with the impedance of CA (i.e., $V_\mathrm{DD,CA} \propto |y_\mathrm{L}|$) in order to approximately maintain a voltage saturation of CA at the first efficiency peak. More specifically, the reconfiguration of $V_\mathrm{DD,CA}$ can be written as
 \begin{equation}
 V_\mathrm{DD,CA}=V_\mathrm{DD,CA0}\sqrt{\frac{1}{\mathrm{real}(z_\mathrm{L})}},
\label{eq:peff}
\end{equation}where $V_\mathrm{DD,CA0}$ is the nominal bias voltage at matched load. The above equation indicates that the reconfigurable DC bias voltage of CA is only related to the real part of load impedance, $z_\mathrm{L}$. For high-resistance loads (i.e., $\mathrm{real}(z_\mathrm{L})>1$), $V_\mathrm{DD,CA}$ should be lower than the nominal value ($V_\mathrm{DD,CA0}$) in order to maintain a constant saturation power of CA. Vice versa, $V_\mathrm{DD,CA}$ should
be increased from $V_\mathrm{DD,CA0}$ for low-resistance loads (i.e., $\mathrm{real}(z_\mathrm{L})<1$). With this load-dependent $V_\mathrm{DD}$ tuning on CA, the saturation power of CA can be well maintained at LBO with a consistent first efficiency peak against arbitrary load mismatch. The variable $V_\mathrm{DD,CA}$ can be realistically supplied by a DC-DC converter against arbitrary $Z_\mathrm{L}$.

\item \textbf{\textit{Doherty Region} ($P_\mathrm{Max}/\mathrm{LBO}\leq P_\mathrm{OUT}<P_\mathrm{Max}/\mathrm{HBO}$)}: when power increases to the Doherty region, the CA should remain Voltage-saturated during load modulation, and ideally BA1 is turned on to load modulate with CA. Under load mismatch, it is found that we can turn on either BA1 or BA2 at LBO to cooperate with CA to form the DPA. Similar to our previous study on PD-LMBA under load mismatch \cite{1d-pdlmba}, it is found that the load dependence of BA$1$’s load modulation (when $\phi=0$) behavior can only be derived by treating CA as a voltage source given by
\begin{equation}
    z_\mathrm{BA1}=\frac{\sqrt{2}V_\mathrm{DD,CA}}{Z_{0}I_{b1}}z_\mathrm{L}+2z_\mathrm{L}
    \label{eq:Z_BA1_Vc}
\end{equation}
\begin{equation}
z_\mathrm{CA}=\frac{V_\mathrm{DD,CA}}{V_\mathrm{DD,CA}z_\mathrm{L}+\sqrt{2}z_\mathrm{L}I_{b1}}.
\label{eq:Z_CA_Vc_b1}
\end{equation}where we can see that the impedance of BA1 is proportional to $Z_\mathrm{L}$. Therefore, when the magnitude of normalized impedance of load is lower than 1 (i.e., $\lvert z_\mathrm{l} \rvert<1$), BA1 should work as the primary peaking  amplifier with BA2 turned off in Doherty region in order to avoid the voltage clipping during load modulation. Likewise, the load dependence of BA2 can only be obtained by treating CA as a current source as written as 
\begin{equation}
z_\mathrm{BA2}=\frac{\sqrt{2}I_{c}}{I_{b2}}\frac{1}{z_\mathrm{L}}+\frac{2}{z_\mathrm{L}},
\label{eq:Z_BA2_IC}
\end{equation}

\begin{equation}
z_\mathrm{CA}=\frac{V_\mathrm{DD,CA}}{V_\mathrm{DD,CA}z_\mathrm{L}-\sqrt{2}I_{b2}}.
\label{eq:Z_CA_Vc_b2}
\end{equation} It can be seen that the impedance of BA2 is inversely proportional to normalized $z_\mathrm{L}$. As a result, when the magnitude of $z_\mathrm{L}$ is greater than 1 (i.e., $\lvert z_\mathrm{L} \rvert>1$), we should exchange the role of BA1 by setting BA2 as the primary peaking amplifier and BA1 off in the Doherty region. In a word, the turning on sequence of BA1 and BA2 depends on the load condition, so that they alternately operates as the primary peaking with different load conditions. It is also interesting to highlight that, similar to PD-LMBA \cite{1d-pdlmba}, the carrier amplifier CA exhibits a current-source/voltage-source duality to BA1 and BA2.

 \item \textbf{\textit{ALMBA Region}($P_\mathrm{OUT}\geq P_\mathrm{Max}/\mathrm{HBO}$)}:
In the ALMBA region with power increases to HBO, the secondary peaking amplifier is turned on which works in conjunction with the primary peaking. In this case, the CA maintains its Current-source/voltage-source duality, such that BA1 and BA2 have complementary dependence of load as indicated as  
 \begin{equation}
    z_\mathrm{BA1}=\frac{\sqrt{2}V_\mathrm{DD,CA}}{Z_{0}I_{b1}}z_\mathrm{L}+2z_\mathrm{L}-\frac{I_{b2}}{I_{b1}}
    \label{eq:Z_BA1_ALMBA}
\end{equation}
\begin{equation}
z_\mathrm{BA2}=\frac{\sqrt{2}I_{c}}{I_{b2}}\frac{1}{z_\mathrm{L}}+\frac{2}{z_\mathrm{L}}-\frac{I_{b1}}{I_{b2}}.
\label{eq:Z_BA2_ALMBA}
\end{equation}

\begin{equation}
z_\mathrm{CA}=\frac{V_\mathrm{DD,CA}}{V_\mathrm{DD,CA}z_\mathrm{L}-\sqrt{2}({I_{b2}-z_\mathrm{L}I_{b1})}}
\label{eq:Z_CA_almba}
\end{equation}
Therefore, BA1 and BA2 can complement each other for power generation under load mismatch just like a standard balanced amplifier. As a result, no reconfiguration is needed in this ALMBA region, and the overall PA performance can be intrinsically sustained against arbitrary load mismatch due to the balanced nature. 
\end{enumerate}

To sum up, a mismatch-resilient three-way load modulation PA can be formed through adjusting the DC voltage of CA and the turning-on sequence of BA1 and BA2 in order to sustain the PA performance in terms of power generation, efficiency and linearity under arbitrary load.

\begin{figure}[t]
\centering
\includegraphics[width=88
mm]{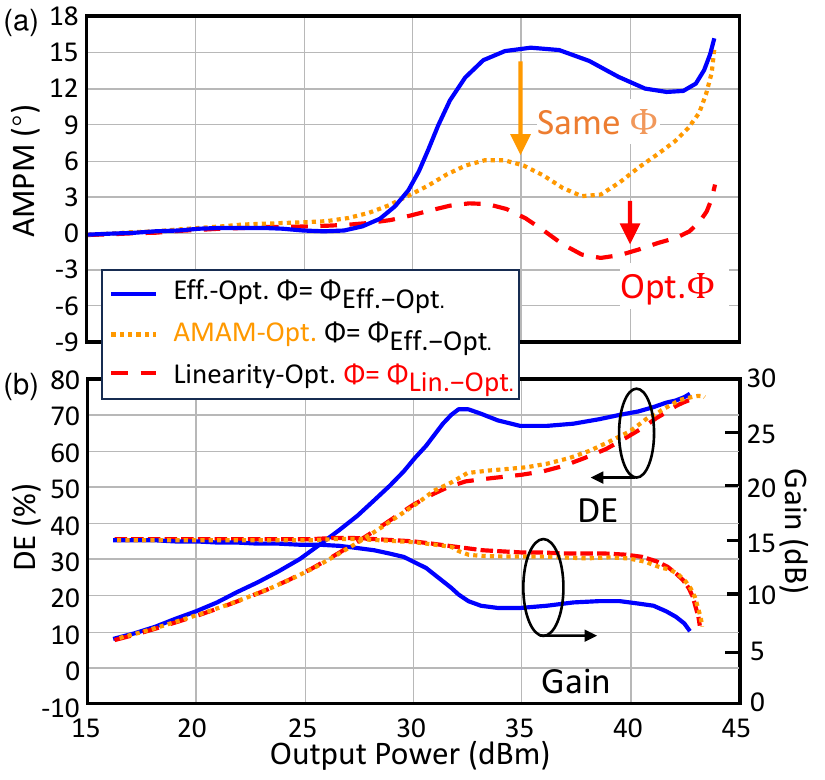}
\vspace{-24pt}
\caption{Optimized performance comparison of linearized H-ALMBA with optimal $\phi$.}
\label{fig8:linerize_performance}
\vspace{-10pt}
\end{figure}
\section{Verification using Emulated H-ALMBA Model }
\label{sec:Verification}

\begin{figure}[t]
\centering
\includegraphics[width=85mm]{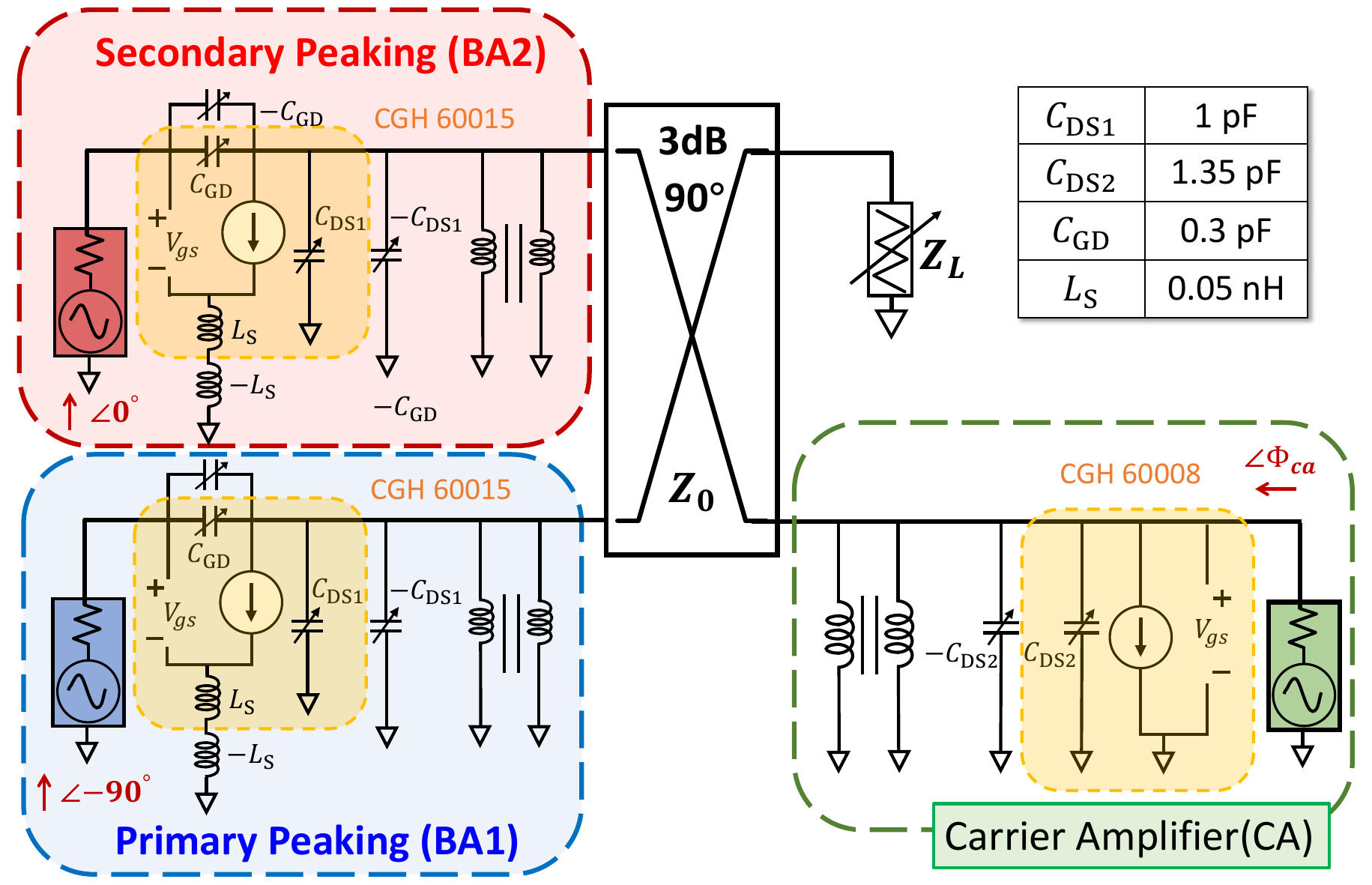}
\vspace{-5pt}
\caption{Emulated model setup of the proposed H-ALMBA with GaN transistors for verification.
}
\label{fig7:Emulated model}
\vspace{-16pt}
\end{figure}
To validate the H-ALMBA theory outlined in Sec.~\ref{sec:Theory_Norm}, an ideal H-ALMBA model using is emulated using two distinct GaN transistor types, ideal quadrature couplers and  transformers for impedance tuning, as depicted in Fig.~\ref{fig7:Emulated model}. Specifically,  CGH60015 model from Wolfspeed  is implemented for BA1 and BA2, and a smaller CGH60008 model is used for CA, given its reduced output power compared to BA. By employing specially designed negative capacitance ($-C_\mathrm{DS}$, $-C_\mathrm{GD}$) and negative source-degeneration inductance ($-L_\mathrm{S}$), the intrinsic parasitic elements of the transistors is fully de-embedded. As a result, the emulated models closely represent ideal transistor devices as voltage-controlled current sources. The complete H-ALMBA emulation model is then created and tested in ADS.

\begin{figure*}
\centering
\includegraphics[width=18cm]{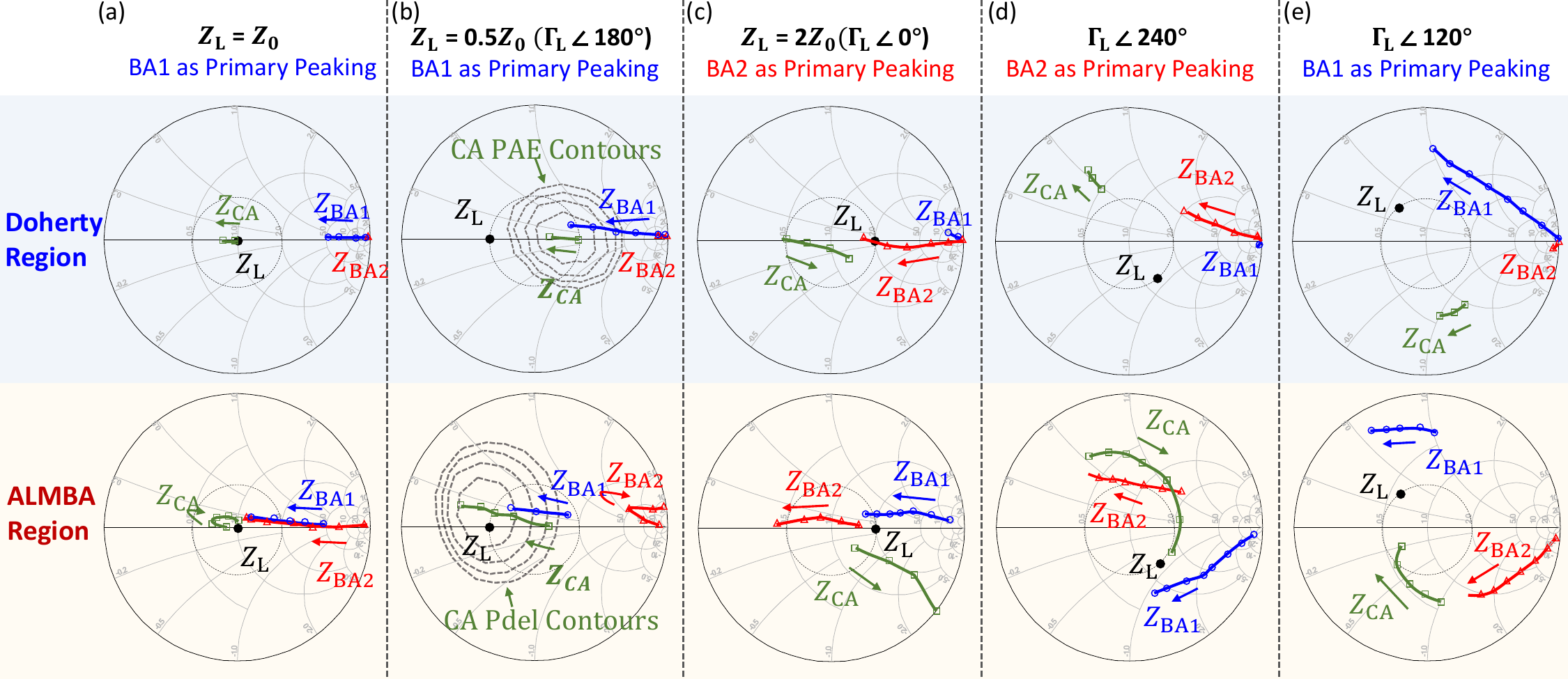}
\vspace{-5pt}
\caption{Load modulation analysis of H-ALMBA for representative cases on VSWR $2:1$ circle in different power regions. (a) $Z_\mathrm{L}=Z_\mathrm{0}$ (b) $\lvert\Gamma_\mathrm{L}\rvert\angle180^{\circ}$ (c) $\lvert\Gamma_\mathrm{L}\rvert\angle0^{\circ}$ (d) $\lvert\Gamma_\mathrm{L}\rvert\angle240^{\circ}$ (e) $\lvert\Gamma_\mathrm{L}\rvert\angle120^{\circ}$.}
\label{fig9:Load_Modulation}
\vspace{-13pt}
\end{figure*}

\vspace{-9pt}
\subsection{High-Linearity Operational Mode of H-ALMBA}
\label{subsec:Emu_lp_region}
Fig.~\ref{fig8:linerize_performance} shows the efficiency and linearity profiles of H-ALMBA towards different design optimums. The standard H-ALMBA design \cite{CM_LMBA} is optimized for efficiency where the first efficiency peak is ensured by the saturation
of CA leading to a sustained high efficiency throughout the entire OBO range. However, strong distortions with $>5$-dB and $>15^\circ$ distortions in AMAM and AMPM, respectively, are observed. Nevertheless, through the linearization techniques introduced in Sec. II-A, the AMAM and AMPM responses can be greatly flattened. Specifically, the AMPM response can mainly be linearized through a linearity-optimal phase-offset which differs from the efficiency-optimal condition, as depicted in Fig.~\ref{fig8:linerize_performance}(a). The AMAM of low-power and Doherty regions can be linearized by slightly lowering the turning-on threshold of BA1, i.e., early turning-on before CA compresses (with slightly compromised efficiency), and the ALMBA region can also be linearized by properly setting the BA2 threshold, as depicted in Fig.\ref{fig8:linerize_performance}(b).


\vspace{-9pt}
\subsection{Reconfiguration for Load-Mismatch Resilience}
\label{subsec:Emu_hp_region}
Note that the low-power region of H-ALMBA is same as PD-LMBA, in which the CA saturation can be maintained at LBO with load-dependent $V_\mathrm{DD,CA}$ as verified in \cite{1d-pdlmba}. Therefore, the verification using emulated model is focused on Doherty and ALMBA regions. The load-modulation behaviors of three sub-amplifiers in H-ALMBA under matched load are illustrated in Fig.~\ref{fig9:Load_Modulation}(a). In this nominal condition, when CA reaches to its saturation, BA1 and BA2 turns on sequentially as primary and secondary peaking amplifiers, and the impedances of BA1 and BA2 are modulated from open circuit to $R_\mathrm{OPT}$ along the real axis on Smith chart. 

Under mismatched load, the roles of BA1 and BA2 are alternated depending on the load condition together with the tuning of $V_\mathrm{DD,CA}$, in order to maintain a load-insensitive H-ALMBA performance. To verify the proposed theory, the high-resistance and low-resistance loads on the $2:1$ VSWR circle are representatively selected to elaborate the load modulation behaviors of BA1, BA2 and CA under different mismatch conditions in Doherty and ALMBA regions.

\begin{enumerate}
    \item \textbf{\textit{High-Resistance Load} ($\lvert z_\mathrm{l} \rvert>1$)}: 
   When the load is mismatched to a high impedance, Fig.~\ref{fig9:Load_Modulation}(c) and (d) shows the load modulation trajectories in pure resistive and complex condition, respectively. In these cases as described in Subsec.~\ref{subsec:theory_mismatch}, BA2 should be set as the primary peaking amplifier that turns on at LBO. In Doherty region, BA1 should remain off in this region with BA2 cooperating with CA. As BA2 has a $1/z_\mathrm{L}$ dependence (see Eq.~\eqref{eq:Z_BA2_IC}), the impedance of BA2 is properly modulated to from open-circuit to a lower value as the power increases, while the CA impedance increases according to Eq.~\eqref{eq:Z_CA_Vc_b2}. When power increases to the ALMBA region, BA1 starts to turn on and be load modulated. It is important to note that the load trajectories of BA1 and BA2 have complementary ending points due to the balanced nature, as indicated by \eqref{eq:Z_BA1_ALMBA} and \eqref{eq:Z_BA2_ALMBA} and proven in Fig.~\ref{fig9:Load_Modulation}(c) and (d).

    \item \textbf{\textit{Low-Resistance Load} ($\lvert z_\mathrm{l} \rvert<1$)}:
Likewise, when load is mismatched to a low impedance, the sub-amplifiers' load modulation trajectories is plotted in Fig.~\ref{fig9:Load_Modulation}(b) and (e). In this case, BA1 maintains the role of matched condition and works as the primary peaking amplifier that turns on at LBO. In Doherty region, BA1 works with CA as a Doherty PA with BA2 remaining off. As BA1 has a $z_\mathrm{L}$ dependence (see \eqref{eq:Z_BA1_Vc}), the impedance of BA1 is properly modulated from open-circuit to a lower value as the power increases, while the CA impedance decreases according to \eqref{eq:Z_CA_Vc_b1}. When the power increases to ALMBA region, the CA load is further modulated to a lower region towards the optimal point of maximum output power. Similar to the previous mismatch-tolerant PD-LMBA theory\cite{1d-pdlmba}, CA substitutes the role of BA2 in power generation, in which the sharp increase of CA current leads to the increase of $Z_\mathrm{BA2}$ based on \eqref{eq:BA2_almba} as illustrated in Fig.~\ref{fig9:Load_Modulation}(b). Nevertheless, the load modulation of BA1 is unaffected such that the overall power generation and dynamic response of H-ALMBA can remain unchanged in this ALMBA region.

\item \textbf{\textit{High VSWR} ($>2:1$)}: In the case where VSWR exceeds $2:1$, the load modulation behavior of the sub-amplifiers shall display a trend similar to that illustrated in Fig.~\ref{fig9:Load_Modulation}, although with a larger scale. Therefore, the proposed reconfiguration method can still be implemented in order to prevent PA from clipping and over-saturation caused by load mismatch, while striving to maximize the power generation and maintain the linearity. However, the peak power and efficiency can inevitably experience a larger degradation due to the increased reflection at higher VSWR. For extreme cases of VSWR $\geq10:1$, the proposed reconfiguration can help protect the devices from breakdown.

\end{enumerate}

\begin{figure}
\centering
\includegraphics[width=9cm]{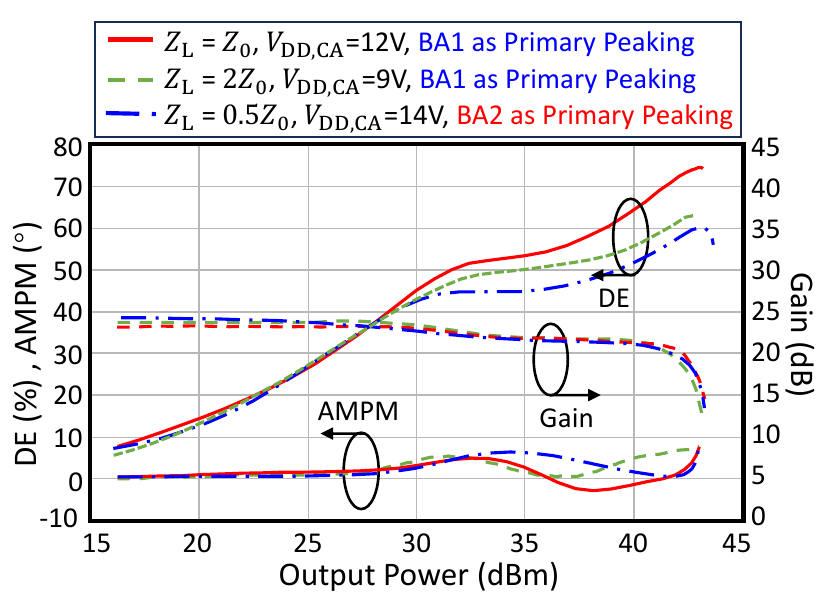}
\vspace{-25pt}
\caption{Simulated efficiency and linearity profile versus $P_\mathrm{OUT}$ of emulated H-ALMBA model under different load conditions with proposed reconfiguration method.}
\label{fig10:EM_simulation_result}
\vspace{-15pt}
\end{figure}

Fig.~\ref{fig10:EM_simulation_result} shows the overall efficiency and linearity profiles of different load conditions. With the tuning of CA supply voltage, $V_\mathrm{CA}$ and the turning on sequence of BA1 and BA2, the first efficiency peak can be kept constantly around $10$-dB OBO, and the linearity response remains on par with that observed under matched conditions. This consistent performance thereby robustly substantiates the theory outlined in Sec.~\ref{sec:Theory_Norm}.

\section{Practical Design of Linear Wideband H-ALMBA with Mismatch Resilience}
\label{sec:Prototype}

Based on the proposed theory, a RF-input linear H-ALMBA is designed with a target frequency range from $1.7-2.9$ GHz. The practical design of this prototype closely aligns with our previous work on hybrid asymmetrical LMBA presented in \cite{CM_LMBA}, while the circuit is re-optimized for high-linearity operation. Specifically, three $10$-W GaN HEMTs (CG2H40010F) from Wolfspeed are used for both CA and BA. In order to accommodate the high PAPR of emerging 5G signals, a back-off range of $10$-dB is targeted for this prototype. Further elaboration on the design intricacies of BA and CA, along with the methodologies for achieving broad bandwidth phase alignment between BA and CA, is expounded in the subsequent subsections.

\begin{figure}[t]
\centering
\includegraphics[width=8.7cm]{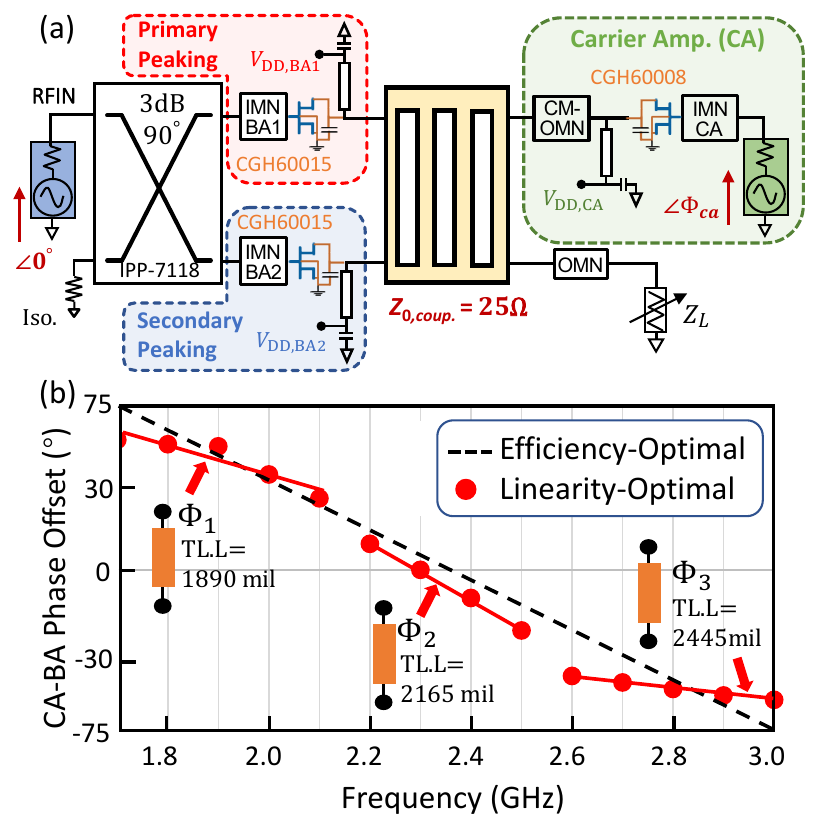}
\vspace{-10pt}
\caption{Design of reconfigurable H-ALMBA: (a) circuit schematic, (b) determined linearity-optimal CA-BA phase offset with corresponding trimmable transmission lane at different frequencies.}
\label{fig11:Paractical design}
\vspace{-11pt}
\end{figure}

\begin{figure}
\centering
\includegraphics[width=8.5cm]{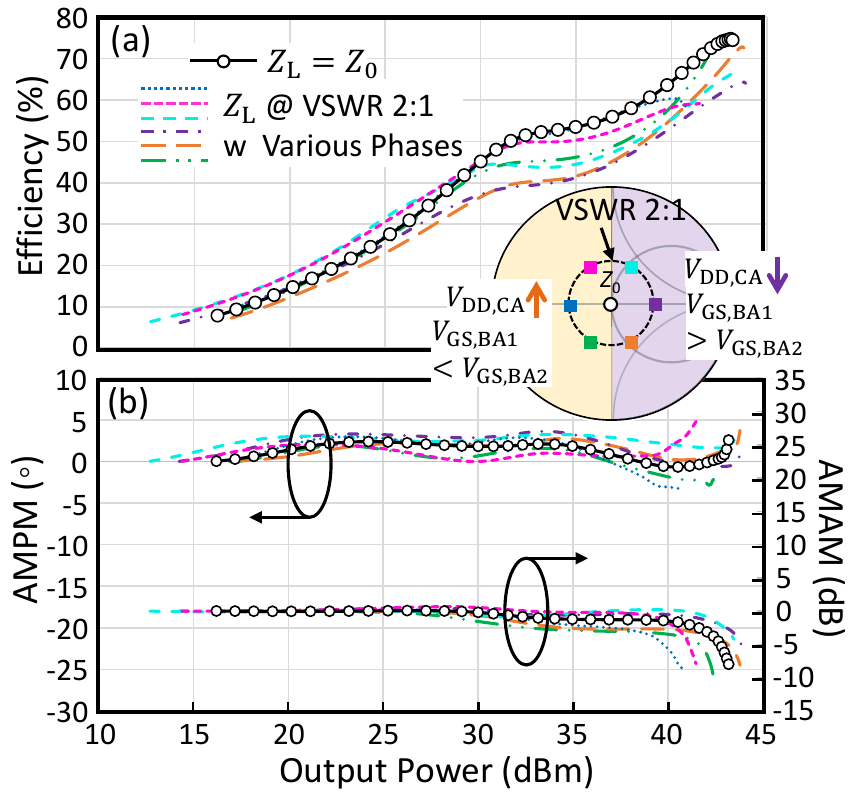}
\vspace{-10pt}
\caption{Simulated H-ALMBA performance with $z_\mathrm{L}$-optimal bias setting under various mismatch conditions: (a) efficiency, (b) linearity.}
\label{fig12:overall_simulation_result}
\vspace{-15pt}
\end{figure}

\vspace{-12pt}
\subsection{Wideband CA and BA Design in Continuous Mode}
Using the sophisticated design method in \cite{CM_LMBA} and \cite{1d-pdlmba}. The CA and BA are both designed with the 10-W GaN transistor (Wolfspeed CG2H40010). Accordingly, the drain of CA is biased in Class-AB mode with a low supply voltage around $10$-V and the matching network is implemented using a simplified TL-based OMN \cite{Chen:Class-E_TMTT2012} to meet the requirement the of the wideband CA design with the highest-possible first efficiency peak. In order to build the balanced amplifier, the input coupler is realized using a commercial component (IPP-7118, available from Innovative Power Products) with a wide bandwidth from $1.7$$-$$2.9$ GHz. The output coupler of BA is implemented with a non-$50$-$\Omega$ three-stage branch-line hybrid structure \cite{Branch-Line_Couplers}, which can provide a sufficient bandwidth and also serve as a part of output impedance matching network (OMN). The OMN of BA is realized by leveraging the input impedance of transformer coupler to set the real part of load admittance and bias line to set the imaginary part. This OMN structure with minimized complexity leads to a low broadband phase dispersion that normally occurs in the complex matching network \cite {LMBA_MWCL2016}, so as to ease the phase equalization between BA and CA in the broadband design. 
\vspace{-12pt}

\begin{figure*}[t]
\centering
\includegraphics[width=17cm]{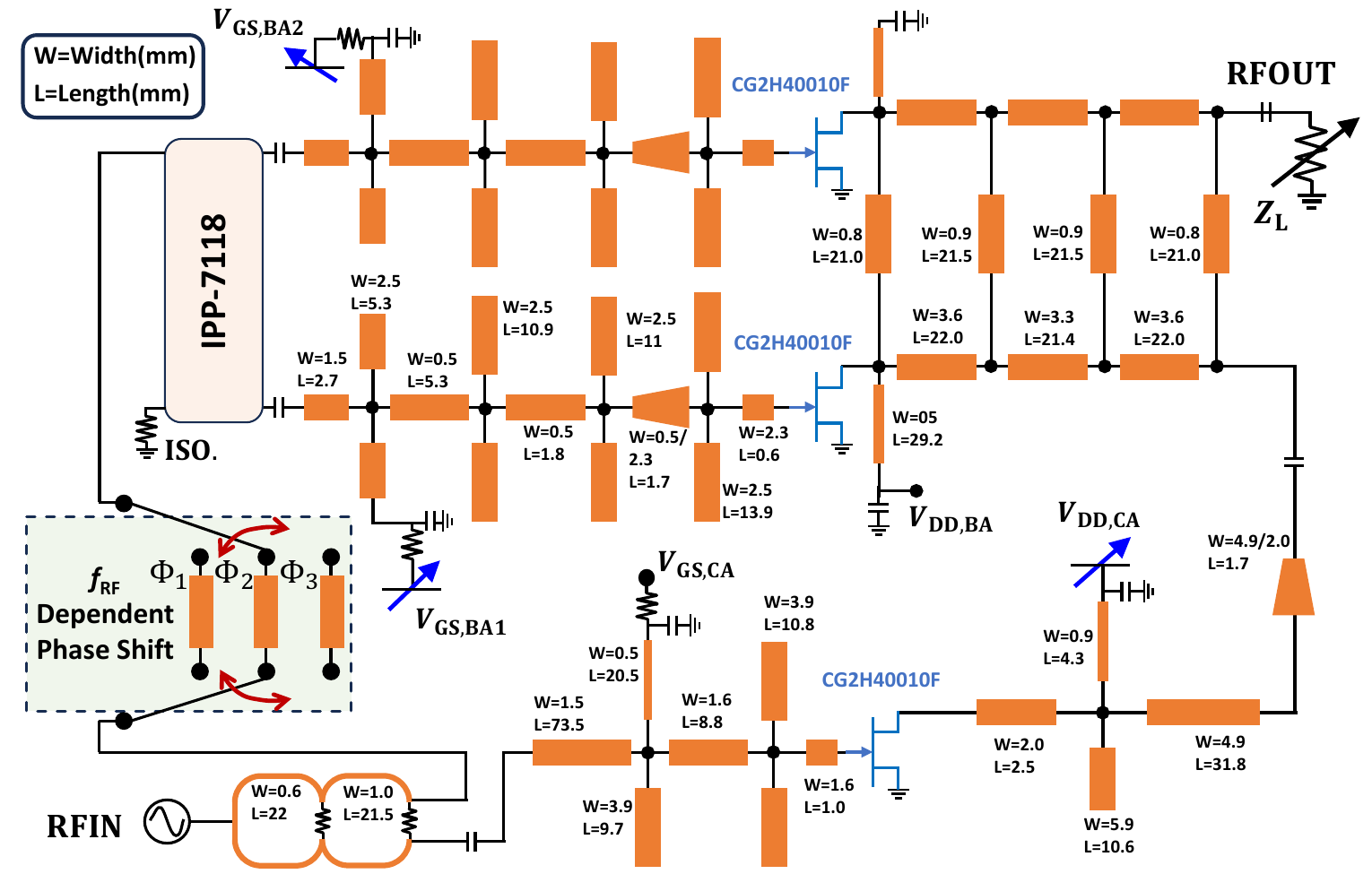}
\vspace{-10pt}
\caption{Circuit schematic overview of designed reconfigurable H-ALMBA.}
\label{fig13:circuit}
\vspace{-15pt}
\end{figure*}

\subsection{Linearity-Optimal BA-CA Phase Offset Design}
To ensure the best efficiency and maximum output power, the optimal phase alignment between CA and BA is required in standard H-ALMBA since the load modulation between the two sub-amplifiers are mainly depend on the CA-BA phase offset. This phase difference can be analysed by setting a phase alignment network introduced in \cite{CM_LMBA}, such that the phase difference between CA and BA can be represent by $\phi_\mathrm{CA}$ as plotted in Fig.~\ref{fig11:Paractical design}(a). In order to find the optimal value of $\phi_\mathrm{CA}$, a control signal is implemented to the isolation port of the BA output quadrature coupler. After a quantity sweeping of $\phi_\mathrm{CA}$, the optimal frequency-dependant phase offset can be found as shown in the dash line of Fig.~\ref{fig11:Paractical design}(b), the ideal phase difference between BA and CA exhibits a nearly linear relationship with frequency, characterized by a negative slope. This phase alignment can be physically realized by using one single electrical length $50-\Omega$ TL fitting the frequency related phase offset\cite{PDLMBA,LMBA_RPC,Dual-Octave-Bandwidth}. The observed TL phase shifter provides a close-to-optimal phase configuration across various frequencies. However, due to the bandwidth limitation of the quadrature coupler and nonlinear dispersion of transistor parasitics, the phase response is no longer pure linear to frequency.

\parskip=0pt In this work, the same phase alignment network is implemented to optimize the CA-BA phase offset targeting the best linearity profile. It is found that the linearity-optimal phase offset between CA and BA is not linearly proportional to frequency as the red dot shown in Fig.~\ref{fig11:Paractical design}(b), which cannot be simply realized using one single transmission line as the efficiency optimal case. However, the phase offset response can be partially linear to the in-band frequency, which separate the phase value into three frequency regions as the red line plotted in Fig.~\ref{fig11:Paractical design}(b). It can be seen that, by using three different electrical length $50$-$\Omega$ TLs, the optimal CA-BA phase offset can be fully covered for the designed bandwidth. In this design, an adjustable phase alignment kit is implemented in realistic measurement, the actual TL length is indicated in Fig.~\ref{fig11:Paractical design}(b). In realistic implementation, this kit can be realized using tunable phase-shifting TL, such as \cite{MMWAVESilicon_WANG2019,LMBA_RPC}. 

The complete circuit design, including a detailed schematic and the specific values for each circuit element, is depicted in Fig.~\ref{fig13:circuit}. With the detailed design described in this section, the overall efficiency and linearity performance at different mismatched load conditions are swept with power as shown in Fig.~\ref{fig12:overall_simulation_result}.

\section{Fabrication and Measurement Results}
\begin{figure}[t]
\centering
 \includegraphics[width=8.9cm]{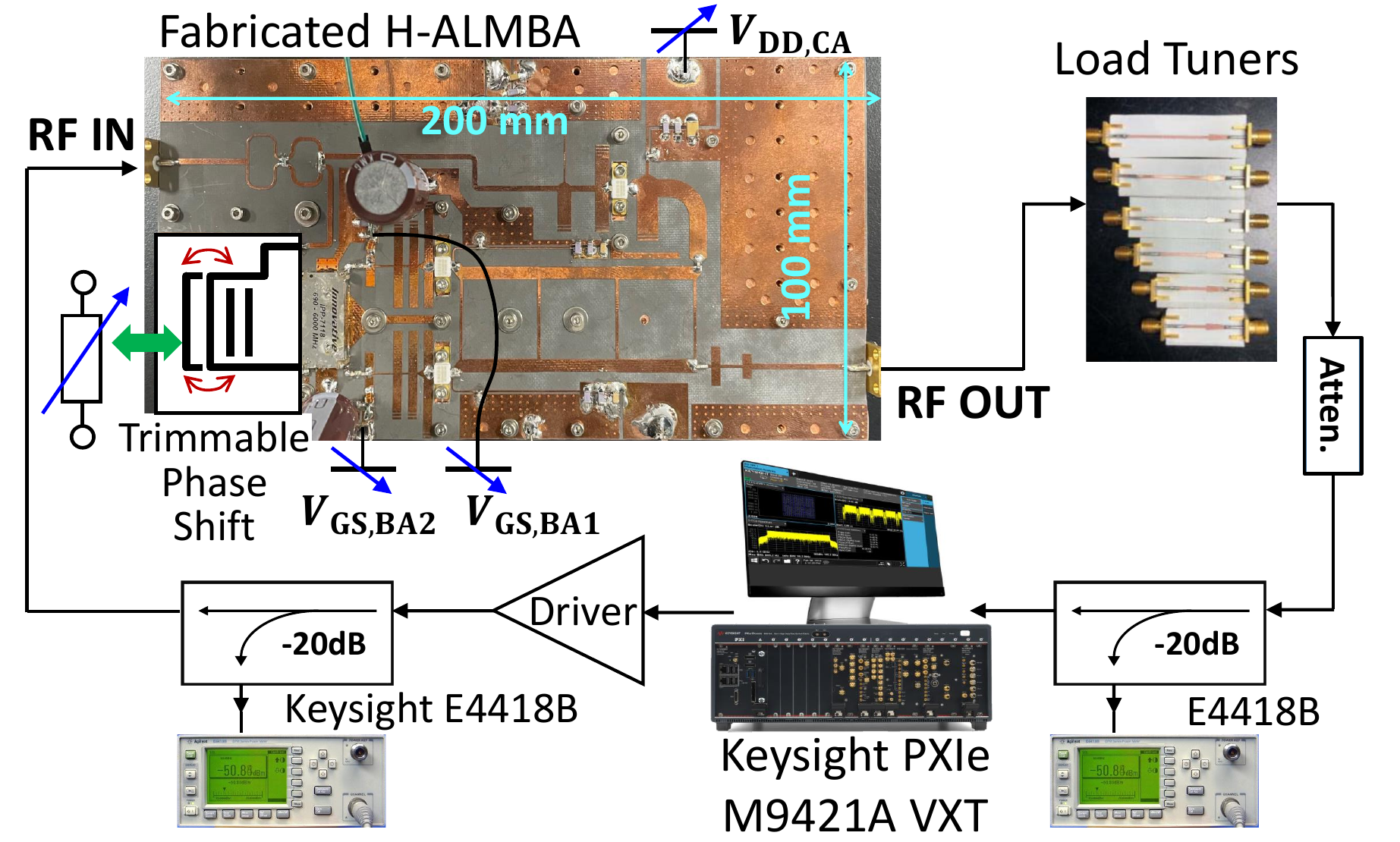}
\vspace{-18pt}
\caption{Fabricated PD-LMBA prototype and testing setup.}
\label{fig14:test_setup}
\vspace{-8pt}
\end{figure}

The measurement system setup including the photograph of fabricated H-ALMBA is shown in  Fig.~\ref{fig14:test_setup}. The board is finalized using a $0.5$-mm ($20$-mil) thick Rogers Duroid-5880 PCB board with a dielectric constant of $2.2$. Both the signal generator and analyzer are using the  Keysight PXIe vector transceiver (VXT M9421) for CW and modulated measurement. In order to realize the mismatch measurement, the circuit output is connected directly to a set of load tuner, which covers the $2:1$ VSWR circle on Smith Chart with the phase swept at $30^{\circ}$ step. Moreover, the frequency-depend phase shifter is realized by several replaceable transmission line which cover the bandwidth. The BAs are biased in Class-C mode with $Z_\mathrm{L}$-dependence $V_\mathrm{GS,BA1}$ and $V_\mathrm{GS,BA2}$. The drain voltage biasing of BA, $V_\mathrm{DD,BA}$, is set to $28$ V to achieve it maximum output power. The CA is biased in Class-F/F$^{-1}$ mode with a $V_\mathrm{GS,CA}$ of $-2.8$ V. In order to maintain its $10$-dB power back-off, the $V_\mathrm{DD,CA}$ is adjustable based on arbitrary load when mismatch.

\begin{figure}[t]
\centering
\includegraphics[width=8.5cm]{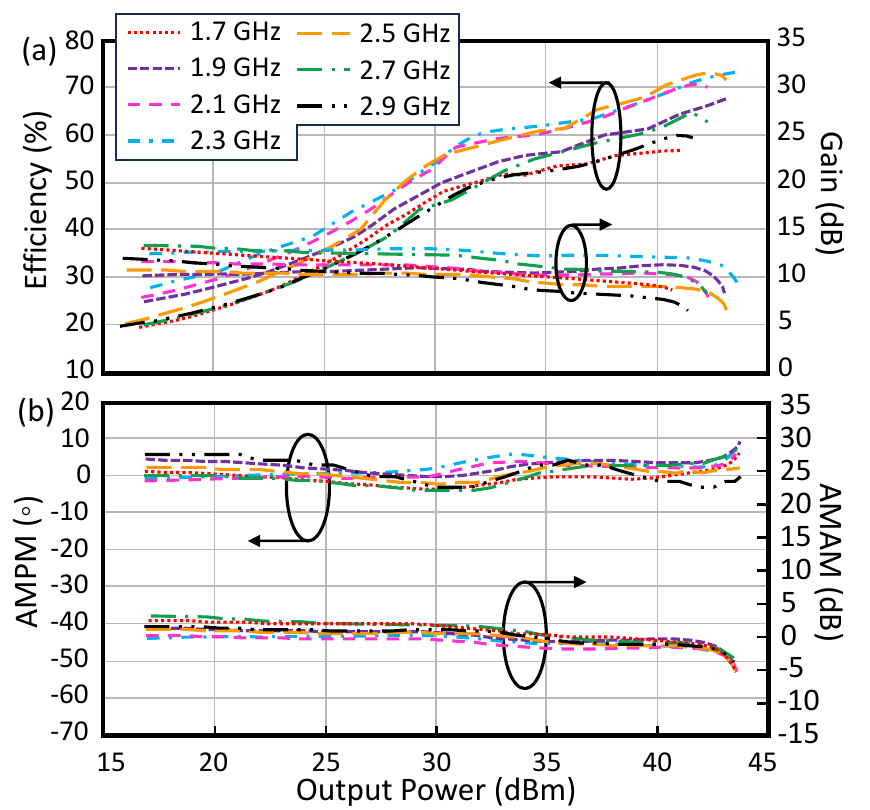}
\vspace{-10pt}
\caption{Measured performance with CW signal under match condition: (a) Efficiency, (b) AMPM and AMAM.}
\label{fig15:cw_match}
\vspace{-12pt}
\end{figure}

\begin{figure}
\centering
\includegraphics[width=8.2cm]{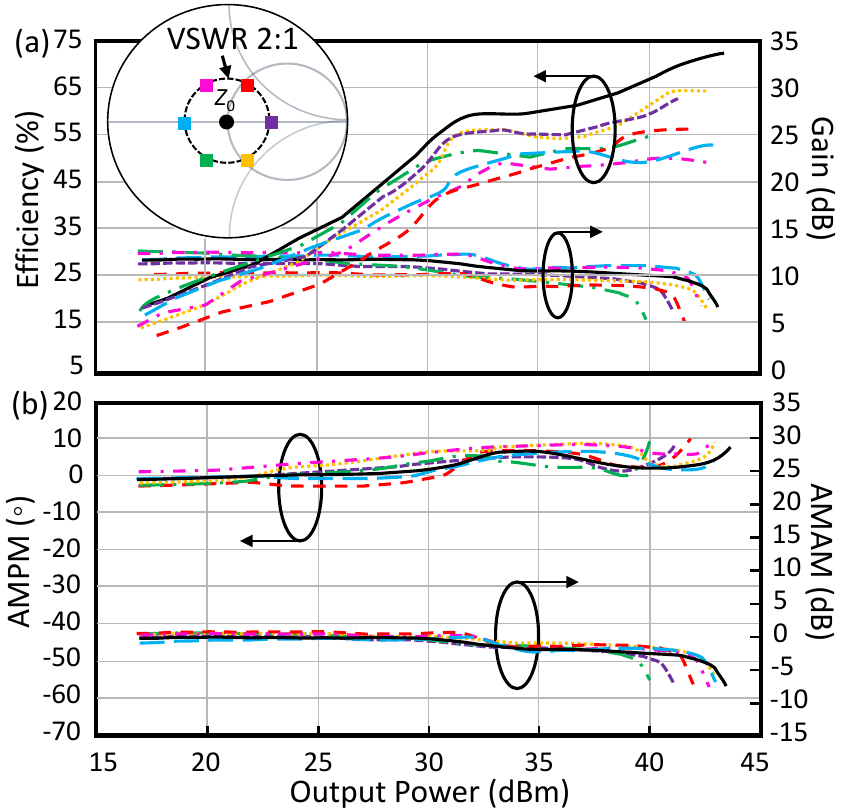}
\vspace{-10pt}
\caption{CW measured performance at $2.2$ GHz for different load conditions on $2:1$VSWR circle in inset Smith: (a) Efficiency, (b) AMPM and AMAM.}
\label{fig16:cwmismatch}
\vspace{-10pt}
\end{figure}

\vspace{-12pt}
\subsection{Continuous-Wave Measurement}
\begin{figure}[t]
\centering
\includegraphics[width=6.8cm]{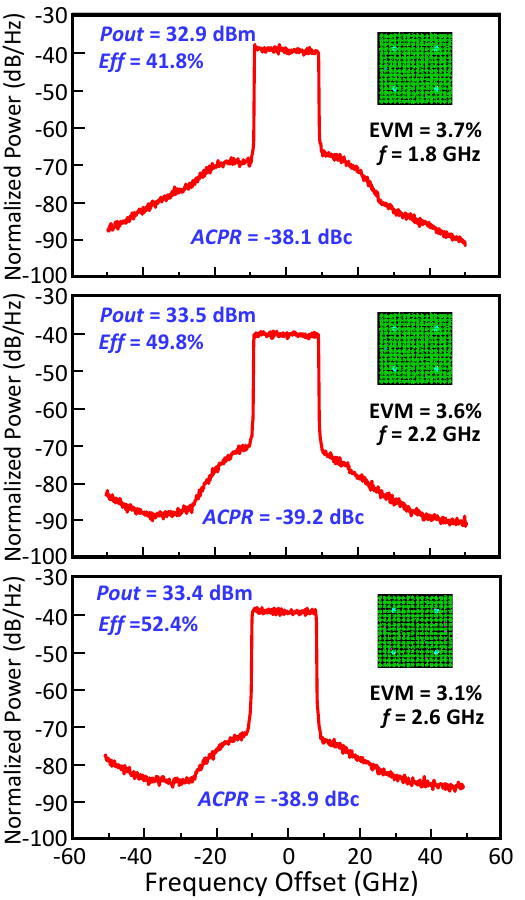}
\vspace{-10pt}
\caption{Modulated measurement results using LTE signal under match condition at different frequencies: (a) 1.8 GHz, (b) 2.2 GHz, (c) 2.6 GHz.}
\label{fig17:mod_match}
\vspace{-18pt}
\end{figure}    

The developed prototype is initially measured using a single-tone continuous-wave (CW) stimulus signal with a frequency range from $1.7$ to $2.9$ GHz at different input power levels. Under a standard matched load impedance of $50$ $\Omega$, the supply voltage $V_\mathrm{DD,CA}$ is set at approximately $10$V, with a slight adjustment to maintain the $10$dB output back-off (OBO) level. The obtained results, encompassing measured efficiency and linearity profile in relation to output power, are graphically depicted in Figure~\ref{fig15:cw_match}. The proposed linear H-ALMBA with an optimal phase and gate biasing achieves a drain efficiency of $56.8\%$$-$$72.9\%$ at peak power and  $49.8\%$$-$$61.2\%$ at $10$-dB OBO with the saturated output power of $41.5$$-$$43.4$ dBm at different frequencies.from $1.7$$-$$2.9$ GHz. The linearity profile of H-ALMBA is flattened with a distortion of AMAM $<7.5$-dB and AMPM $<8.5^\circ$.

In order to verify the mismatch recovery capability of H-ALMBA through proposed reconfiguration, the CW measurement with various mismatched loads is further evaluated and a set of load tuners are fabricated as a customized load-pull system. At $2.2$ GHz, the measured DE and linearity profile versus swept output power at various loads across the $2:1$ VSWR circle are shown in Fig.~\ref{fig16:cwmismatch}. With the proposed $Z_\mathrm{L}$-optimal biasing, a DE from $47.1\%$ to $64.8\%$ is measured at peak power under various load mismatch conditions, and the measured DE at $10$-dB OBO is up to $57.2\%$. It is clearly seen that the efficiency profiles under load mismatch well replicate the shape at matched condition. Moreover, the linearity profile is well maintained under mismatched condition with a distortion of AMAM $<8.9$-dB and AMPM $<9.6^\circ$.

\vspace{-12pt}
\subsection{Modulated Measurement}
\label{sec:modulated}

In order to validate the linearity of the proposed circuit under realistic communication condition, the modulated measurement is presented with a $20$ MHz modulation-bandwidth single-carrier $64$ QAM LTE signal at three in-band frequency $1.7$, $2.1$ and $2.5$ GHz. The measurement is first performed under matched condition. Fig.~\ref{fig17:mod_match} shows the the measured power spectral density (PSD) under matched condition, the ACPR of most measured frequencies are higher than 20 dB without any digital pre-distortion. Fig.~\ref{fig17:mod_match}
shows the proposed H-ALMBA achieves the average drain efficiency (DE) of $41.8\%$$-$$52.4\%$ and average output power of $32.9$$-$$33.5$ dBm over the designed frequency band. 
          
\begin{figure}
\centering
\includegraphics[width=8.8cm]{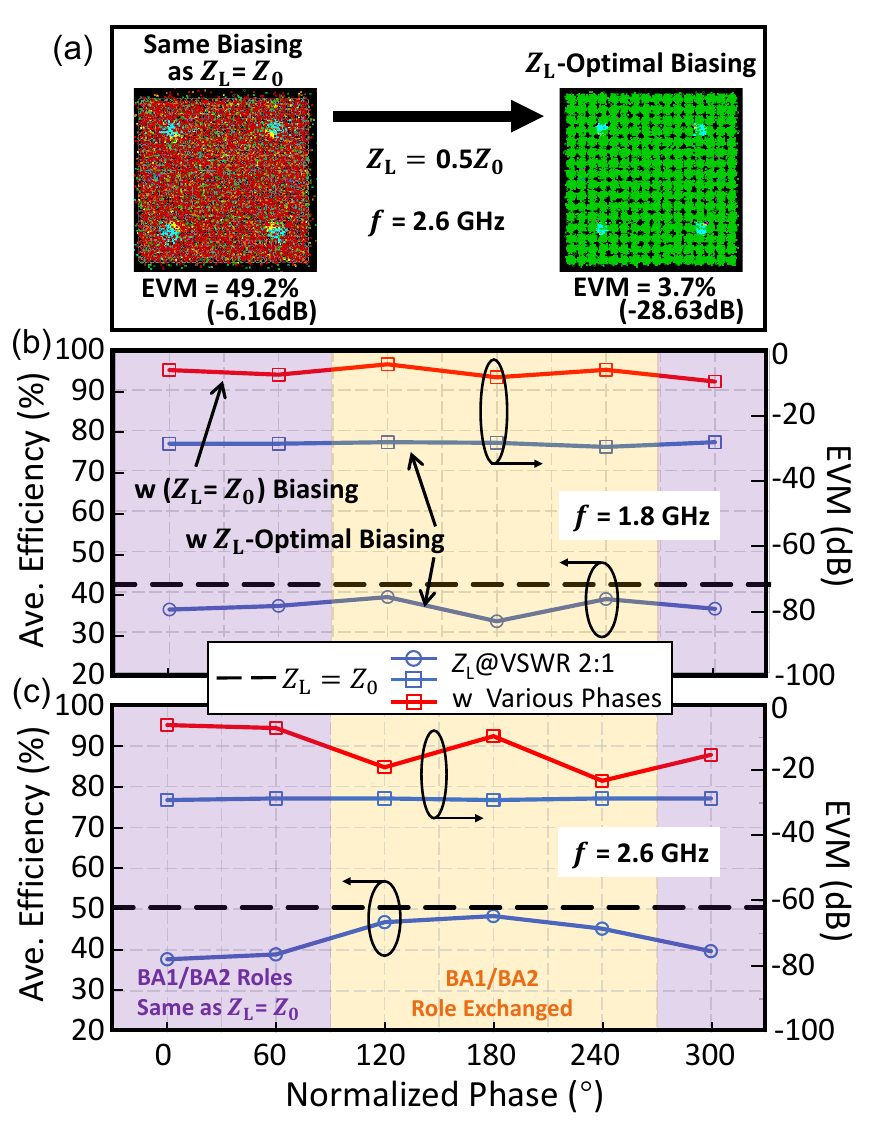}
\vspace{-25pt}
\caption{Modulation measurements results across 2 : 1 VSWR at 1.8 GHz and 2.6 GHz with comparison between different biasing modes.}
\label{fig18:Mod_mis}
\vspace{-15pt}
\end{figure}    
The proposed H-ALMBA is further measured under the $2$:$1$ VSWR load mismatch to present the mismatch-resilient feature. Fig.~\ref{fig18:Mod_mis} shows the optimal results
obtained with six phase mismatch load over VSWR 2:1 circle, it achieves average drain efficiency of $44.7\%$$-$$53.5\%$ with the average output power of  $29.8\%$$-$$30.6\%$ dBm at $1.7$ GHz, average drain efficiency of $52.1\%$$-$$59.1\%$ with the average output power of $30$$-$$31.3$ dBm at $2.1$ GHz and average drain efficiency of $45.5\%$-$55\%$ with the average output power of $29$$-$$30.8$ dBm at $2.5$ GHz.

Additionally, it is found in measurement that the linearity of H-ALMBA is devastated with load mismatch, i.e., EVM collapsed to $~50\%$ as the sample point shown in Fig.~\ref{fig18:Mod_mis}(a). Nevertheless, the EVM can be perfectly recovered through the proposed biasing reconfiguration. This $Z_\mathrm{L}$-dependent reconfiguration is conducted over the entire $2:1$ VSWR circle as shown in Fig.~\ref{fig18:Mod_mis}(b) and (c). A low EVM lower to $3.6\%$ and average efficiency up to $48.9\%$ can both be experimentally maintained against various load conditions at two represented frequencies.

A comprehensive comparison between the proposed linear H-ALMBA and other recently reported mismatch-resilient PAs is shown in Table \uppercase\expandafter{\romannumeral1}. It is important to emphasize that the first ever linear reconfiguration operation under mismatch is accomplished with a very competitive profile. Specifically, an overall wider operation bandwidth is implemented in either match or mismatch condition. Therefore, The proposed H-ALMBA not only clearly exhibits better mismatch resilience but also exhibit excellent linearity profile under both match and mismatch condition which well validates the effectiveness of the proposed theory. 
 
\newcommand{\tabincell}[2]{\begin{tabular}{@{}#1@{}}#2\end{tabular}}
\begin{table*}
\footnotesize
\renewcommand{\arraystretch}{1.2}
\label{tab:state}
\caption{\small Comparison with State-of-the-Art of Recently Reported Mismatch-Resilient PAs}
\end{table*}

\begin{figure*}
\centering
\vspace{-20pt}
\includegraphics[width=\textwidth]{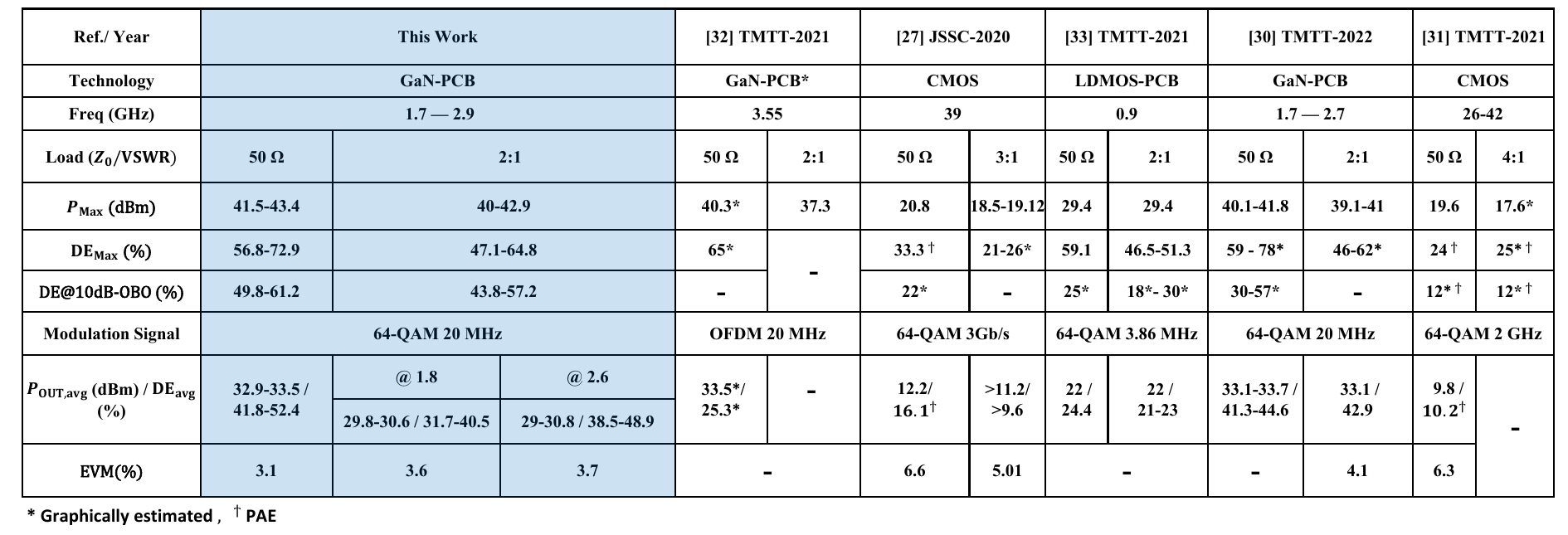}
\label{fig:state}
\vspace{-35pt}
\end{figure*}

\vspace{-8pt}
\section{Conclusion}

The paper presents, for the first time, a wideband highly linear and load-insensitive three-way load-modulation power amplifier based on H-ALMBA topology. It includes an control amplifier (CA) and two peaking balanced amplifiers (BA1 and BA2), achieving high-order load modulation over a wide bandwidth. Through rigorous analysis and derivation, it is found that the turning-on sequence of peaking devices BA1 and BA2 can be reconfigured based on varying load conditions, addressing their load impedance-dependent complementary nature. This reconfiguration helps mitigate non-linearity issues caused by CA overdrive. Note that the reconfigurable biasing of PAs has been widely applied in realistic systems, e.g., the multi-band multi-mode PAs in mobile handsets. More importantly, this reconfiguration distinguishes itself by its simplicity, eliminating the need for a dual input\cite{dual_input_pedro}. Additionally, AMPM distortion is corrected by adjusting the phase offset between the CA and BA, in tandem with the CA's Current Source (CS) - Voltage Source (VS) duality. This adjustment is further enhanced by setting the load-dependent voltage ($V_\mathrm{DD,CA}$). As a result, this design is resilient to load mismatch and adaptable to different frequencies, demonstrating significant efficiency and linearity improvements under various load conditions. To empirically substantiate this theory, a prototype is created, and its measurements reveals high linearity across its bandwidth and high efficiency at both $10$-dB back-off and peak power. Additionally, it effectively handles modulated signal transmission at in-band frequencies under both matched load and  $2$:$1$ VSWR load mismatch. This technology demonstrates considerable promise for massive MIMO Power Amplifier applications in future wireless communication systems.


%



\ifCLASSOPTIONcaptionsoff
  \newpage
\fi

\bibliographystyle{ieeetr}
\bibliography{References.bib}









\end{document}